# Virtual birefringence imaging and histological staining of amyloid deposits in label-free tissue using autofluorescence microscopy and deep learning

## Authors

Xilin Yang[1,2,3], Bijie Bai[1,2,3], Yijie Zhang[1,2,3], Musa Aydin[1,4], Sahan Yoruc Selcuk[1,2,3], Zhen Guo[1], Gregory A. Fishbein[5], Karine Atlan[6], William Dean Wallace[7], Nir Pillar[1,2,3*], Aydogan Ozcan[1,2,3,8*]

## Affiliations

[1]Electrical and Computer Engineering Department, University of California, Los Angeles, CA, 90095, USA.

[2]Bioengineering Department, University of California, Los Angeles, 90095, USA.

[3]California NanoSystems Institute (CNSI), University of California, Los Angeles, CA, USA.

[4]Department of Computer Engineering, Fatih Sultan Mehmet Vakif University, Istanbul, Turkiye

[5]Department of Pathology and Laboratory Medicine, David Geffen School of Medicine at the University of California, Los Angeles, CA, 90095, USA.

[6]Department of Pathology, Hadassah Hebrew University Medical Center, Jerusalem, Israel

[7]Department of Pathology, Keck School of Medicine, University of Southern California, Los Angeles, CA, 90033, USA.

[8]Department of Surgery, University of California, Los Angeles, CA, 90095, USA.

[*]npillar@g.ucla.edu

[*]ozcan@ucla.edu

## Abstract

Systemic amyloidosis is a group of diseases characterized by the deposition of misfolded proteins in various organs and tissues, leading to progressive organ dysfunction and failure. Congo red stain is the gold standard chemical stain for the visualization of amyloid deposits in tissue sections, as it forms complexes with the misfolded proteins and shows a birefringence pattern under polarized light microscopy. However, Congo red staining is tedious and costly to perform, and prone to false diagnoses due to variations in the amount of amyloid, staining quality and expert interpretation through manual examination of tissue under a polarization microscope. Here, we report the first demonstration of virtual birefringence imaging and virtual Congo red staining of label-free human tissue to show that a single trained neural network can rapidly transform autofluorescence images of label-free tissue sections into brightfield and polarized light microscopy equivalent images, matching the histochemically stained versions of the same samples. We demonstrate the efficacy of our method with blind testing and pathologist evaluations on cardiac tissue where the virtually stained images agreed well with the histochemically stained ground truth images. Our virtually stained polarization and brightfield images highlight amyloid birefringence patterns in a consistent, reproducible manner while mitigating diagnostic challenges due to variations in the quality of chemical staining and manual imaging processes as part of the clinical workflow.

Introduction

Systemic amyloidosis is a heterogeneous group of disorders characterized by the deposition of abnormally folded proteins in tissue. The clinical picture of systemic amyloidosis is not specific, with profound fatigue, weight loss, and edema being the common presenting symptoms[1]. The real prevalence of systemic amyloidosis is not known[2,3]. A retrospective evaluation of kidney biopsies suggests that amyloidosis is not as rare as it is thought to be, accounting for ~43% of nephrotic proteinuria above age 60[4]. In another study, 31% of the patients with multiple myeloma had confirmed evidence of systemic amyloidosis[5]. Cardiac amyloid deposition, causing infiltrative/restrictive cardiomyopathy, is the leading cause of morbidity and mortality in systemic amyloidosis, regardless of the underlying pathogenesis of amyloid production[6]. Similar to other organs involved with amyloidosis, cardiac amyloidosis remains substantially underdiagnosed, and it is advised to test for cardiac amyloidosis presence during the initial work-up of all patients ⩾65 years old hospitalized with heart failure[7]. Early diagnosis of systemic amyloidosis is essential to reducing morbidity and mortality of the disease. A prompt intervention following early-stage amyloid detection may save patients from extensive and irreversible tissue damage. In addition, a definitive diagnosis has become increasingly important since a number of impactful treatment options have developed[8].

Diagnosis of amyloidosis is usually based on the demonstration of amyloid deposits in a tissue biopsy. Cardiac biopsy provides the most definitive diagnostic evidence in amyloid cardiomyopathy, and endomyocardial biopsy was shown to be a safe and relatively simple procedure[9]. Congo red is considered the gold standard stain used in the vast majority of histopathology laboratories to identify amyloid in tissues. When tested under cross-polarized light microscopy, Congo red-stained amyloid areas demonstrate birefringence, which is considered a specific feature of amyloidosis. However, the traditional workflow (as depicted in Figure 1a) has several drawbacks. Congo red staining analysis exhibits inter-observer variability, partly attributed to the challenging nature of Congo red staining[10]. In addition to a standard brightfield microscopy evaluation, visualization under polarized light microscopy is needed to examine the presence of birefringence. The quality of the examining microscope for highlighting amyloid birefringence can significantly limit the accuracy of pathologist evaluations, leading to an increase in both false-negative and false-positive results. A false negative tissue pathology report can mislead diagnosticians and lead to the exclusion of the amyloidosis diagnosis, often without reconsideration among the differential diagnoses. Recent reports showed that the median time from the symptom onset to amyloidosis diagnosis was ~2 years. Additionally, nearly a third of patients reported seeing at least 5 physicians before receiving a diagnosis of amyloidosis[11].

These technical challenges of Congo red staining and diagnostic inspection under polarized light microscopy also pose a barrier to the broader adoption of digital pathology[12,13]. There is currently no clinically approved digital pathology slide scanner with the required polarization imaging components, and the existing scanning polarization imagers are limited to well-resourced settings, mostly for research use[14].

Recently, deep learning-based technologies have introduced transformative opportunities to biomedical research and clinical diagnostics[13,15] using deep neural networks (DNNs) to learn intrinsic structures in large datasets[16]. A notable application of this is the use of deep learning for virtual histological staining of label-free tissue samples[17–26]. In this technique, a deep convolutional neural network (CNN) is trained to computationally stain microscopic images of unstained (label-free) tissue sections, matching their histologically stained counterparts. This computational approach aims to circumvent some of the challenges associated with traditional histochemical staining. Multiple research groups and institutions explored this deep learning-based virtual staining technique and successfully utilized it for digitally generating a wide range of routinely used stains, both histological[17,27–29], immunohistochemical[20,30] and fluorescent stains[31]. Such deep-learning-based transformations can also be applied from one stain type to another or to blend several stains simultaneously[19,32–34], utilizing existing stained tissue images to provide more information that could help with diagnosis.

Here, we report a deep learning-based virtual tissue staining technique designed to digitally label amyloid deposits in unstained, label-free tissue sections (as illustrated in Figure 1). This technique, for the first time, achieves autofluorescence to birefringence image transformations along with virtual Congo red staining of label-free tissue. It employs conditional generative adversarial networks[35] (cGANs) to rapidly and digitally convert autofluorescence microscopy images of unstained tissue slides into virtually stained birefringence and brightfield images, closely resembling the corresponding images of the histochemically stained samples, helping the identification of amyloid deposits in label-free tissue slides. We designed our deep learning model to simultaneously learn the cross-modality image transformations for the two output modalities within a single neural network, performing both *autofluorescence-to-birefringence* and *autofluorescence-to-brightfield* image transformations using a digital staining matrix (Figure 1b), which is an additional channel concatenated to the input autofluorescence microscopy images, indicating the desired output modality (birefringence vs. brightfield). An auxiliary registration module was also integrated into the cGAN in the training process to mitigate spatial misalignments in the training data by learning to register the output and the ground truth histochemically stained images[36] (Figure 5a). After a one-time training phase, when given a new label-free test sample, the virtual staining network successfully generated microscopic images of the Congo red-stained tissue in both brightfield and birefringence channels. These digitally generated, virtually stained images were then stitched into whole slide images (WSI) and examined by pathologists using a customized multi-modality WSI viewer, offering the flexibility to swiftly toggle between the brightfield and polarization/birefringence views. Our methodology's effectiveness was affirmed by three board-certified pathologists attesting to the non-inferior quality of the images generated by our network model compared to histochemically stained images, demonstrating a high degree of concordance with the ground truth. Validated by a group of pathologists, our approach effectively bypasses the limitations of traditional histological staining workflow manually performed by histotechnologists and also eliminates the need for tedious polarization imaging with specialized optical components, facilitating a faster and more reliable diagnosis of amyloidosis.

## Results

### Label-free virtual birefringence imaging and amyloid staining

We demonstrated our virtual birefringence imaging and virtual Congo red staining technique by training a deep-learning model on a dataset comprising label-free tissue sections with a total of 386 and 65 training/validation and testing image patches, respectively, where each image patch had 2048×2048 pixels, all obtained from eight distinct patients. To obtain this dataset, we imaged unlabeled/label-free tissue sections with a 4-channel autofluorescence microscope and then sent these label-free slides for standard histochemical Congo red staining (for ground truth generation). The histochemically stained tissue slides were then imaged using a standard brightfield microscope scanner as well as a polarization microscope (see the Methods section). Manual fine-tuning of the polarizer and analyzer settings was conducted by a board-certified pathologist to ensure the quality of the captured birefringence patterns that served as our ground truth for the polarization channel. Following the image data acquisition, an image registration process was performed in the training phase to spatially register the brightfield and the polarization images to label-free autofluorescence images to mitigate potential image misalignments[19,37]. The total data size for all the training, validation and testing images amounts to ~40 GB. For more details on image dataset acquisition and preprocessing of data, refer to the Method section.

Our deep learning model was composed of three sub-modules: (1) a generator designed to learn the two necessary cross-modality image transformations, i.e., autofluorescence-to-birefringence and autofluorescence-to-brightfield imaging, (2) a discriminator that engages in adversarial learning to differentiate between the output and ground truth images, thereby aiding the generator during the training process, and (3) an image registration module tasked with aligning the output images with the ground truth, which helps to mitigate residual misalignments within the dataset. The image transformations from autofluorescence to brightfield and birefringence modalities were learned within a single neural network model. To determine which output (brightfield or birefringence) is desired, a digital staining matrix was concatenated to the input. This digital staining matrix, matching the pixel count and shape of the input images, determines the output modality on a per-pixel basis with "1" indicating *brightfield* and "-1" indicating *polarization/birefringence* channel (Figure 1b). During the training, we mixed the target images from both modalities and concatenated the corresponding digital staining matrix to the autofluorescence input images.

After the model convergence, we conducted a blind evaluation of our model using 65 test images (each with 2048×2048 pixels) obtained from two previously unseen patients. During the model testing phase, brightfield and birefringence output images were naturally aligned with the corresponding input autofluorescence images. For both the brightfield and polarization channels, the predicted virtual images exhibited a high degree of agreement with the ground truth images, as demonstrated in Fig. 2, which showcases side-by-side visual comparisons between the virtually stained images produced by our deep learning model alongside the corresponding histochemically stained ground-truth images. This figure includes two representative slides from distinct patients. Specifically, in Figure 2(a), the right side displays zoomed-in sections of the histochemically stained brightfield images, revealing regions of interest (ROIs) with a pinkish hue indicative of congophilic areas. In the corresponding polarization images, these regions exhibit an apple-green birefringence, characteristic of amyloid deposits. The virtually created images of these same regions also manifest the same morphology of amyloid presence, affirming that our inference model has effectively learned to transform autofluorescence label-free images into both brightfield and birefringence images, matching the histochemically stained counterparts, presenting an accurate appearance of amyloid deposits. In Figure 2(b), a similar comparative analysis is shown for another patient. In this case, the histochemically stained images display a specific area, labeled as ROI3, which is an area without amyloid deposits ("negative region"). No congophilic features can be identified in the brightfield image, and no apple-green birefringence is seen in the polarization channel. Our virtual staining model correctly replicated these morphological characteristics, generating images that align closely with the histochemically stained ground truth images, without introducing false positive staining.

## Pathologist evaluations and performance quantification

To further assess the effectiveness of our virtual staining approach, three board-certified pathologists were engaged to blindly evaluate the quality of histochemically stained and virtually stained images. The evaluation comprised two distinct parts: 1) assessing the image quality of brightfield Congo-red stain and 2) evaluating the appearance and quality of amyloid deposits using large, bundled patch images of the polarization and brightfield channels. The first part involved examining small patches, solely from brightfield modality, which were randomly cropped from histochemically stained and virtually stained images without overlapping FOVs (see Supplementary Figure 1a). A total of 163 test image patches (each with $1024 \times 1024$ pixels) underwent random augmentations, were shuffled and then presented to the pathologists in a blinded manner. The expert evaluations concentrated on four primary metrics: stain quality of nuclei (M1), cytoplasm (M2), and extracellular space (M3), as well as the staining contrast of congophilic areas (M4). The first three metrics (M1-M3) are standard for evaluating stained tissue images, while the last one (M4) is unique to Congo red staining, clinically relevant for amyloidosis diagnosis. For all metrics, the grading scale ranged from 1 to 4, where 4 represents "perfect", 3 represents "very good", 2 represents "acceptable", and 1 represents "unacceptable" quality. Figure 3a visualizes the results using violin plots to compare the scores given to histochemically stained and virtually stained images. The distributions of these plots show no major differences for any of the evaluation metrics. The mean values for each pathologist (P1-P3) are plotted in Figure 3b with an error bar representing the standard deviations (also see Supplementary Table 1). For the first three metrics (M1-M2-M3), slightly higher scores are given to the histochemical images compared to the virtually stained images, with a mean difference of 0.229 (5.72%), 0.334 (8.35%), and 0.114 (2.85%), averaging across all pathologists and all patches - out of a scale of 4. However, for M4, which is the most relevant for Congo red staining, the difference in the mean pathologist scores falls to a negligible level of ~0.067 (1.67%) out of 4. Overall, the stain quality scores corresponding to the virtual and histochemical staining are closely matched, each falling within their respective standard deviations, as shown in Fig. 3. Additional image FOVs supporting this conclusion can be found in Supplementary Figure 2.

The second part of the expert evaluations focused on the analysis of birefringence appearance and image quality using larger FOVs and included bundled brightfield and polarization images. The same FOVs from histochemically stained and virtually stained images were included with different image augmentations (Supplementary Figure 1b). This evaluation incorporated three distinct metrics (M5-M7): amyloid quantification (M5), the quality of birefringence appearance (M6), and the consistency between brightfield and polarization images (M7). The grading scale ranges from 1 to 3 and is different for each metric: percentage for amyloid quantification with one-third spacing (i.e., 1 = 100%); very good, moderate, and non-diagnostic for apple-green birefringence image quality; categories of low, medium, and high for inconsistency between brightfield and polarized Congo red images (lower scores indicate higher quality with less inconsistent features). The pathologists conducted their evaluations using a custom image viewer that offered a user-friendly interface, enabling easy toggling between the brightfield and polarization images of the same FOV and detailed examination of tissue areas of interest (refer to the Methods). Prior to the evaluation, the pathologists were familiarized with the examination process through a brief tutorial that utilized scored examples from the training dataset. Figure 4a presents a violin plot corresponding to our test image set, comparing the distributions of these scores (M5-M7) given to histochemical and virtually stained images. For metrics M6 and M7, the expert score distributions for the virtually stained images surpassed those of the histochemically stained images, with mean improvements of 0.17 (5.67 %) and 0.24 (8.33%) for M6 and M7, respectively. For M5, however, the mean difference in performance dropped to an insignificant level of 0.0066 (0.22%) in favor of the histochemically stained images. Figure 4b further depicts the mean and the standard deviation values across all the samples for each pathologist (also see Supplementary Table 1).

Note that because the histochemically stained and virtually stained images for the same tissue FOVs were both included in the expert evaluation process, we can also compare the pathologists' scores for individual image FOVs. For this image pair-based analysis, we first compared the average scores of the 3 pathologists on the same image FOVs, generated for the histochemically stained and virtually stained images. The results of this paired analysis are displayed in Figure 4c, where each column contains three metrics for a given FOV. A yellow-colored entry represents a higher score given to the virtually stained image output for that tissue FOV, while a green color for an entry represents that a lower score is given to the virtually stained image FOV compared to the histochemically stained counterpart; finally, a grey color represents equal scores. These analyses reported in Fig. 4c reveal that the virtually stained label-free images were blindly given higher or equal performance scores by the expert panel in 66%, 91%, and 89% of the tissue FOVs for metrics M5, M6 and M7, respectively. A more detailed comparison of each pathologist's individual scores is also reported in Supplementary Figure 4. Some randomly selected tissue FOVs with performance scores are also provided in Supplementary Figure 3. These quantitative comparisons further demonstrate the success of our virtual birefringence imaging and label-free tissue staining method using deep learning.

## Discussion

In this manuscript, we trained and tested a virtual birefringence imaging and virtual Congo red staining approach that allows for amyloidosis identification. Our innovative approach utilizes label-free tissue autofluorescence texture to create a virtual Congo red-stained slide with both brightfield and birefringence image images, enabling accurate identification of amyloid deposits within label-free tissue.

In recent years, there have been accelerated efforts to overcome some of the challenges seen in traditional glass slide-based pathology. These have led to the development and adoption of novel imaging systems and WSI scanners that have helped with the transition of pathology into the digital era. While WSI scanners have demonstrated their capabilities to digitize various histochemical and immunohistochemical stained slides, automatically digitizing slides under polarized light remains technically challenging in histology labs due to delicate polarization components and varying illumination conditions. When visualizing a tissue slide under polarized light, constant adjustments of the tissue orientation and polarizer are required to ensure accurate and robust detection of birefringence patterns. In fact, none of the commercially available, clinically approved WSI scanners in the digital pathology field can automatically digitize birefringence images of tissue samples. As a result, pathologists continue to rely on manually operated light microscopes and standard polarizers for amyloid deposit inspection/detection, which is also at the heart of diagnostician-induced errors. Moreover, the quality of the light microscope and polarizers significantly influences the detection ability of amyloidosis by the practicing pathologist. In instances with low amounts of amyloid deposits within the tissue, the microscope's ability to visualize the slide with high contrast and resolution can make the difference between a correct and false diagnosis. By training a virtual staining neural network to transform label-free tissue autofluorescence images into birefringence and brightfield microscopy images that are equivalent to the Congo red stained images of the same tissue, as it would normally appear after chemical staining, we offer a deep learning-based solution to both the challenging nature of the histochemical Congo red staining process as well as the digitization of the birefringence images of stained tissue sections, which currently does not exist in clinical WSI scanners. Our virtual polarization imaging and tissue staining method ensures consistent and reproducible imaging of amyloid deposits within label-free tissue, eliminating the manual processes needed in both the chemical staining of tissue and the constant polarizer and tissue adjustments routinely performed by diagnosticians when using a polarized light microscope. Bypassing such manually operated polarization microscopes, our method can be readily adopted on standard pathology WSI scanners approved for digital pathology by simply using standard filter sets commercially available for capturing autofluorescence images of tissue slices. This also eliminates the need for microscopy hardware changes or specialized optical parts in digital pathology scanners that are already deployed.

Quantitative and comparative analyses of three board-certified pathologists revealed that the virtually stained slides, with their birefringence and brightfield image channels, were on par with chemically stained Congo red slides. Since our label-free approach is not critically dependent on manual labor in its staining and imaging processes, it is particularly beneficial in cases with low amounts of amyloid deposits, which can be easily missed when a diagnostician examines the slide without a high-quality polarizing microscope. In such cases, our method may increase the diagnostic yield and decrease false negative rates. An interesting future direction can be automated label-free detection of regions of amyloid deposits to highlight the problematic areas, which may further assist pathologists with attention heatmaps to accelerate diagnosis speed and reduce false negative rates.

The presented label-free approach has several inherent advantages compared to traditional histochemical Congo red staining. It minimizes staining artifacts compared to chemical staining techniques; it substantially reduces reliance on manual labor, and the virtual staining process diminishes the utilization of hazardous chemicals during slide preparation. It is important to emphasize that all the training processes described for the virtual staining algorithm, including the use of different modalities and processing pipelines, are a one-time effort. Once the model is trained, the blind inference process (virtual staining with brightfield and birefringence channels) of a new, unknown sample will only require seconds per FOV (e.g., the inference time for a FOV size of $2048 \times 2048$ pixels is < 2 sec); therefore, it can transform a label-free whole slide within a few minutes using a state-of-the-art GPU.

In addition to the technical challenges in amyloidosis detection, the spotty nature of the disease and variations in the density of amyloid deposits increase the odds of false diagnoses. Thick tissue sections (e.g., 8-10 µm) are recommended for accurate Congo red birefringence visualization, as they provide more intense staining and allow for the identification of smaller amyloid deposits compared to the commonly used 4 µm sections in pathology. Nonetheless, thicker sections can negatively affect the amount of residual tissue and may have a detrimental impact on the depletion of small biopsy blocks, such as those used in cardiac biopsies. Insufficient tissue is often a limiting factor in performing additional stains or molecular studies. We believe that our method, relying on tissue autofluorescence rather than imaging under polarized light, would be less sensitive to tissue thickness and potentially can assist clinicians in reaching the same diagnostic conclusions while sparing tissue.

In conclusion, we reported the first demonstration of virtual birefringence imaging and virtual tissue staining that effectively transforms label-free cardiac tissue slides into images that match their Congo red-stained histochemical counterparts, viewable in both brightfield and polarization channels. Our method, as a fully digital process, can generate virtually stained WSIs in minutes with high repeatability, eliminating the variations and limitations associated with manual tissue staining methods and the use of polarization microscope components. By reducing the turnaround time, manual labor, and reliance on hazardous chemicals, this method can potentially transform the traditional workflow for the diagnosis of amyloidosis and set the stage for a large-scale, multi-center trial to further validate the clinical utility of our results.

## Methods

### Sample preparation and data acquisition

Unlabeled heart tissue sections were obtained from USC Keck School of Medicine under IRB #HS-20-00151. Following the clinical standard for amyloid inspection, 8 μm sections were cut from archived cardiac tissue blocks that tested positive for Congo red. Samples with less than 5 mm$^2$ tissue area or less than 5% amyloid-involved tissue, determined by the original pathology report, were excluded. After autofluorescence scanning, the standard Congo red staining was performed.

Autofluorescence images of label-free tissue samples were taken using a conventional scanning fluorescence microscope (IX-83, Olympus) equipped with a ×40/0.95NA objective lens (UPLSAPO, Olympus). These images were captured at four distinct excitation and emission wavelengths, each using a fluorescent filter set in a filter cube: DAPI (Semrock DAPI-5060C-OFX, EX 377/50 nm, EM 447/60 nm), FITC (Semrock FITC-2024B-OFX, EX 485/20 nm, EM 522/24 nm), TxRed (Semrock TXRED-4040C-OFX, EX 562/40 nm, EM 624/40 nm), and Cy5 (Semrock CY5-4040C-OFX, EX 628/40 nm, EM 692/40 nm). The autofluorescence images were recorded using a scientific complementary metal-oxide-semiconductor (sCMOS) image sensor (ORCA-flash4.0 V2, Hamamatsu Photonics) using exposure times of 150 ms, 500 ms, 500 ms, and 1000 ms for the DAPI, FITC, TxRed, and Cy5 filters, respectively. μManager (version 1.4) software[38], designed for microscope management, was used for the automated image capture process. Autofocus[39] is applied on the first autofluorescence channel (DAPI) for each FOV. Following the completion of the standard Congo red staining, high-resolution brightfield WSIs were obtained using a scanning microscope (AxioScan Z1, Zeiss) with a ×20/0.8NA objective lens (Plan-Apo) at the Translational Pathology Core Laboratory (TPCL) at UCLA. Polarized images of Congo red-stained slides were captured using a modified conventional brightfield microscope (IX-83, Olympus) with a halogen lamp used as the illumination source without blue filters. The microscope was equipped with a linear polarizer (U-POT, Olympus) and an analyzer with an adjustable wave plate (U-GAN, Olympus) with a ×20/0.75NA objective lens (UPlanSApo). The linear polarizer was placed on the condenser adapter (between the light source and the sample slide) and fixed in a customized 3D printed holder (Ultimaker S3, PETG black) to ensure polarization orientation during the scanning process. The analyzer was inserted into the slider-compatible revolving nosepiece, between the sample slide and the image sensor.

The orientations of all polarization components were adjusted by a board-certified pathologist for optimal image quality.

Image preprocessing and registration

The registration process is essential to successfully train an image-to-image translation network. An alternative is to apply CycleGAN-like architectures[32,40] which do not require paired samples for training. However, such unpaired image-based approaches result in inferior image quality with potential hallucinations compared to training with precisely registered image pairs[17,20,37]. In this work, we registered both the brightfield and birefringence images individually to match the autofluorescence images of the same samples using a two-step registration process. First, we stitched all the images into WSIs for all three imaging modalities and globally registered them by detecting and matching speeded-up robust features (SURF) on downsampled WSIs[41,42]. Then, we estimated the spatial transformations (projective) using the detected features with an M-estimator sample consensus algorithm[43] and accordingly warped the brightfield and birefringence WSIs. Following this, the coarsely matched autofluorescence, brightfield and birefringence WSIs were divided into bundles of image tiles, each consisting of 2048×2048 pixels. Using these image tiles, we further improved the accuracy of our image registration to address optical aberrations among different imaging systems and morphological changes that occurred in the histochemical staining process with a correlation-based elastic registration algorithm[17,44]. During this elastic registration process, a registration neural network model was trained to align the style of the autofluorescence images with the styles of the brightfield and birefringence images. The registration model used for this purpose shared the same architecture and training strategy as our virtual staining network (detailed in the next section), however, without a registration submodule and was only used for the data preparation stage. After the image style transformation using the registration model, the pyramid elastic image registration algorithm was applied. This process involved hierarchically matching the local features of the sub-image blocks of different resolutions and calculating transformation maps. These transformation maps were then used to correct the local distortions in the brightfield and birefringence images, resulting in a better match with their autofluorescence counterparts. This training and registration process was repeated for both brightfield and birefringence images until precise pixel-level registration was achieved. The registration steps were implemented using MATLAB (MathWorks), Fiji[45] and PyTorch[46].

Pathologists' blind evaluations

Three board-certified pathologists were included to blindly evaluate the image quality of histochemically stained and virtually stained images. The blind evaluations were conducted in two ways: (1) Brightfield Congo red stain image quality on small image patches; (2) Large patch bundled images with both brightfield and birefringence images of the same FOV. For part 1 evaluation, small patches ($1024 \times 1024$ pixels, each corresponding to ~$166 \times 166$ μm$^2$) are randomly cropped from histochemical and virtual images without any overlapping FOVs. Then, each image is randomly augmented in the following ways: original, left-right flip, top-bottom flip, and random rotation at 90, 180 and 270 degrees. A total of 163 images were randomly shuffled and sent to pathologists without labeling. Pathologists scored four metrics evaluating the brightfield Congo red image quality: stain quality of nuclei, stain quality of cytoplasm, stain quality of extracellular space and stain contrast of congophilic areas. For part 2, we selected 56 large image patches ($2048 \times 2048$ pixels, each corresponding to ~$332 \times 332$ μm$^2$), and collected the brightfield and birefringence images of the same FOVs. For each image bundle, we collected both histochemically and virtually stained images, and randomly augmented the image bundles while ensuring a different augmentation method for each, resulting in a total of 112 image bundles to be scored. Pathologists scored three metrics evaluating the birefringence channel: amyloid quantification (% of positive areas / total tissue surface), birefringence image quality, and inconsistency between brightfield and polarization Congo red images. For part 2, the evaluation of the bundled images, we customized the Napari (0.4.18) viewer[55] as a graphical user interface (GUI) for pathologists to conveniently evaluate the high-resolution images. The interface features keyboard bindings for switching between brightfield and birefringence views, selecting images, and controlling zoom functions for regions of interest using a mouse. All the participating pathologists were given a short tutorial on using the GUI, which included scored example FOVs selected from the training dataset. In both evaluations, any images that pathologists declined to give a score are excluded from the analysis.

## Supplementary Information

- Network architecture and training strategy
- Supplementary Figure 1. Data preparation and the workflow of pathologist evaluations.
- Supplementary Figure 2. Examples of image patches and pathologist scores for bright-field Congo red stain image quality.
- Supplementary Figure 3. Examples of larger image patches and pathologist scores for birefringence image quality.
- Supplementary Figure 4. Confusion matrices for the birefringence image quality scores of pathologists (P1, P2 and P3) blindly comparing virtually stained and histochemically stained images.
- Supplementary Table 1. Average values and standard deviations of the pathologists' evaluation scores.

## References

1. Baker, K. R. & Rice, L. The amyloidoses: clinical features, diagnosis and treatment. *Methodist DeBakey Cardiovasc. J.* **8**, 3–7 (2012).

2. Picken, M. M. The Pathology of Amyloidosis in Classification: A Review. *Acta Haematol.* **143**, 322–334 (2020).

3. Amyloidosis - Statistics. *Cancer.Net* https://www.cancer.net/cancer-types/amyloidosis/statistics (2012).

4. Maixnerova, D. *et al.* Nationwide biopsy survey of renal diseases in the Czech Republic during the years 1994-2011. *J. Nephrol.* **28**, 39–49 (2015).

5. Desikan, K. R. *et al.* Incidence and impact of light chain associated (AL) amyloidosis on the prognosis of patients with multiple myeloma treated with autologous transplantation. *Leuk. Lymphoma* **27**, 315–319 (1997).

6. Gertz, M. A., Lacy, M. Q. & Dispenzieri, A. Amyloidosis: recognition, confirmation, prognosis, and therapy. *Mayo Clin. Proc.* **74**, 490–494 (1999).

7. Gilstrap, L. G. *et al.* Epidemiology of Cardiac Amyloidosis–Associated Heart Failure Hospitalizations Among Fee-for-Service Medicare Beneficiaries in the United States. *Circ. Heart Fail.* **12**, e005407 (2019).

8. Benson, M. D. *et al.* Tissue biopsy for the diagnosis of amyloidosis: experience from some centres. *Amyloid* **29**, 8–13 (2022).

9. Martinez-Naharro, A., Hawkins, P. N. & Fontana, M. Cardiac amyloidosis. *Clin. Med. Lond. Engl.* **18**, s30–s35 (2018).

10. Klunk, W. E., Pettegrew, J. W. & Abraham, D. J. Quantitative evaluation of congo red binding to amyloid-like proteins with a beta-pleated sheet conformation. *J. Histochem. Cytochem.* **37**, 1273–1281 (1989).

11. Maurer, M. S., Elliott, P., Comenzo, R., Semigran, M. & Rapezzi, C. Addressing Common Questions Encountered in the Diagnosis and Management of Cardiac Amyloidosis. *Circulation* **135**, 1357–1377 (2017).

12. Al-Janabi, S., Huisman, A. & Van Diest, P. J. Digital pathology: current status and future perspectives. *Histopathology* **61**, 1–9 (2012).

13. Song, A. H. *et al.* Artificial intelligence for digital and computational pathology. *Nat. Rev. Bioeng.* **1**, 930–949 (2023).

14. Patel, A. *et al.* Contemporary Whole Slide Imaging Devices and Their Applications within the Modern Pathology Department: A Selected Hardware Review. *J. Pathol. Inform.* **12**, 50 (2021).

15. Madabhushi, A. & Lee, G. Image analysis and machine learning in digital pathology: Challenges and opportunities. *Med. Image Anal.* **33**, 170–175 (2016).
11

16. LeCun, Y., Bengio, Y. & Hinton, G. Deep learning. *nature* **521**, 436–444 (2015).

17. Rivenson, Y. *et al.* Virtual histological staining of unlabelled tissue-autofluorescence images via deep learning. *Nat. Biomed. Eng.* **3**, 466–477 (2019).

18. Zhang, Y. *et al.* Digital synthesis of histological stains using micro-structured and multiplexed virtual staining of label-free tissue. *Light Sci. Appl.* **9**, 78 (2020).

19. de Haan, K. *et al.* Deep learning-based transformation of H&E stained tissues into special stains. *Nat. Commun.* **12**, 4884 (2021).

20. Bai, B. *et al.* Label-Free Virtual HER2 Immunohistochemical Staining of Breast Tissue using Deep Learning. *BME Front.* **2022**, (2022).

21. Zhang, Y. *et al.* Virtual Staining of Defocused Autofluorescence Images of Unlabeled Tissue Using Deep Neural Networks. *Intell. Comput.* **2022**, 2022/9818965 (2022).

22. Bai, B. *et al.* Deep learning-enabled virtual histological staining of biological samples. *Light Sci. Appl.* **12**, 57 (2023).

23. Kreiss, L. *et al.* Digital staining in optical microscopy using deep learning - a review. *PhotoniX* **4**, 34 (2023).

24. Cao, R. *et al.* Label-free intraoperative histology of bone tissue via deep-learning-assisted ultraviolet photoacoustic microscopy. *Nat. Biomed. Eng.* **7**, 124–134 (2023).

25. Abraham, T. M. *et al.* Label- and slide-free tissue histology using 3D epi-mode quantitative phase imaging and virtual hematoxylin and eosin staining. *Optica* **10**, 1605–1618 (2023).

26. Pillar, N., Li, Y., Zhang, Y. & Ozcan, A. Virtual Staining of Non-Fixed Tissue Histology. *Mod. Pathol.* 100444 (2024).

27. Mayerich, D. *et al.* Stain-less staining for computed histopathology. *Technology* **3**, 27–31 (2015).

28. Rivenson, Y. *et al.* PhaseStain: the digital staining of label-free quantitative phase microscopy images using deep learning. *Light Sci. Appl.* **8**, 23 (2019).

29. Li, Y. *et al.* Virtual histological staining of unlabeled autopsy tissue. *Nat. Commun.* **15**, 1684 (2024).

30. Cone, B. *et al.* Spatial Overlay: A novel approach for evaluating tumor microenvironment (TME) specific expression of PD-L1 in whole slide images of lung cancer.

31. Patrick, D., Moghtader, J., Wang, H., de Haan, K. & Rivenson, Y. Deep Learning-Enabled Virtual H&E and Fluoro- Jade B Tissue Staining for Neuronal Degeneration.

32. Liu, S. *et al.* Unpaired stain transfer using pathology-consistent constrained generative adversarial networks. *IEEE Trans. Med. Imaging* **40**, 1977–1989 (2021).

33. Ghahremani, P. *et al.* Deep learning-inferred multiplex immunofluorescence for immunohistochemical image quantification. *Nat. Mach. Intell.* **4**, 401–412 (2022).

34. Yang, X. *et al.* Virtual Stain Transfer in Histology via Cascaded Deep Neural Networks. *ACS Photonics* **9**, 3134–3143 (2022).

35. Mirza, M. & Osindero, S. Conditional Generative Adversarial Nets. Preprint at https://doi.org/10.48550/arXiv.1411.1784 (2014).

36. Kong, L., Lian, C., Huang, D., Hu, Y. & Zhou, Q. Breaking the dilemma of medical image-to-image translation. *Adv. Neural Inf. Process. Syst.* **34**, 1964–1978 (2021).

37. Wang, H. *et al.* Deep learning enables cross-modality super-resolution in fluorescence microscopy. *Nat. Methods* **16**, 103–110 (2019).

38. Edelstein, A. D. *et al.* Advanced methods of microscope control using μManager software. *J. Biol. Methods* **1**, e10 (2014).

39. Redondo, R. *et al.* Autofocus evaluation for brightfield microscopy pathology. *J. Biomed. Opt.* **17**, 036008–036008 (2012).

40. Zhu, J.-Y., Park, T., Isola, P. & Efros, A. A. Unpaired image-to-image translation using cycle-consistent adversarial networks. in *Proceedings of the IEEE international conference on computer vision* 2223–2232 (2017).

41. Bay, H., Tuytelaars, T. & Van Gool, L. SURF: Speeded Up Robust Features. in *Computer Vision – ECCV 2006* (eds. Leonardis, A., Bischof, H. & Pinz, A.) vol. 3951 404–417 (Springer Berlin Heidelberg, Berlin, Heidelberg, 2006).

42. Lowe, D. G. Distinctive Image Features from Scale-Invariant Keypoints. *Int. J. Comput. Vis.* **60**, 91–110 (2004).

43. Torr, P. H. & Zisserman, A. MLESAC: A new robust estimator with application to estimating image geometry. *Comput. Vis. Image Underst.* **78**, 138–156 (2000).

44. Saalfeld, S., Fetter, R., Cardona, A. & Tomancak, P. Elastic volume reconstruction from series of ultra-thin microscopy sections. *Nat. Methods* **9**, 717–720 (2012).

45. Schindelin, J. *et al.* Fiji: an open-source platform for biological-image analysis. *Nat. Methods* **9**, 676–682 (2012).

46. Paszke, A. *et al.* Pytorch: An imperative style, high-performance deep learning library. *Adv. Neural Inf. Process. Syst.* **32**, (2019).

47. Goodfellow, I. *et al.* Generative adversarial nets. *Adv. Neural Inf. Process. Syst.* **27**, (2014).

48. Arar, M., Ginger, Y., Danon, D., Bermano, A. H. & Cohen-Or, D. Unsupervised multi-modal image registration via geometry preserving image-to-image translation. in *Proceedings of the IEEE/CVF conference on computer vision and pattern recognition* 13410–13419 (2020).

49. Balakrishnan, G., Zhao, A., Sabuncu, M. R., Guttag, J. & Dalca, A. V. VoxelMorph: a learning framework for deformable medical image registration. *IEEE Trans. Med. Imaging* **38**, 1788–1800 (2019).

50. Huber, P. J. Robust Estimation of a Location Parameter. *Ann. Math. Stat.* **35**, 73–101 (1964).

51. Oktay, O. *et al.* Attention U-Net: Learning Where to Look for the Pancreas. Preprint at https://doi.org/10.48550/arXiv.1804.03999 (2018).

52. Maas, A. L., Hannun, A. Y. & Ng, A. Y. Rectifier nonlinearities improve neural network acoustic models. in *Proc. icml* vol. 30 3 (Atlanta, GA, 2013).

53. He, K., Zhang, X., Ren, S. & Sun, J. Deep residual learning for image recognition. in *Proceedings of the IEEE conference on computer vision and pattern recognition* 770–778 (2016).

54. Kingma, D. P. & Ba, J. Adam: A Method for Stochastic Optimization. Preprint at http://arxiv.org/abs/1412.6980 (2017).

55. Chiu, C.-L., Clack, N. & Community, T. N. napari: a Python Multi-Dimensional Image Viewer Platform for the Research Community. *Microsc. Microanal.* **28**, 1576–1577 (2022).

# Figures and Figure Captions

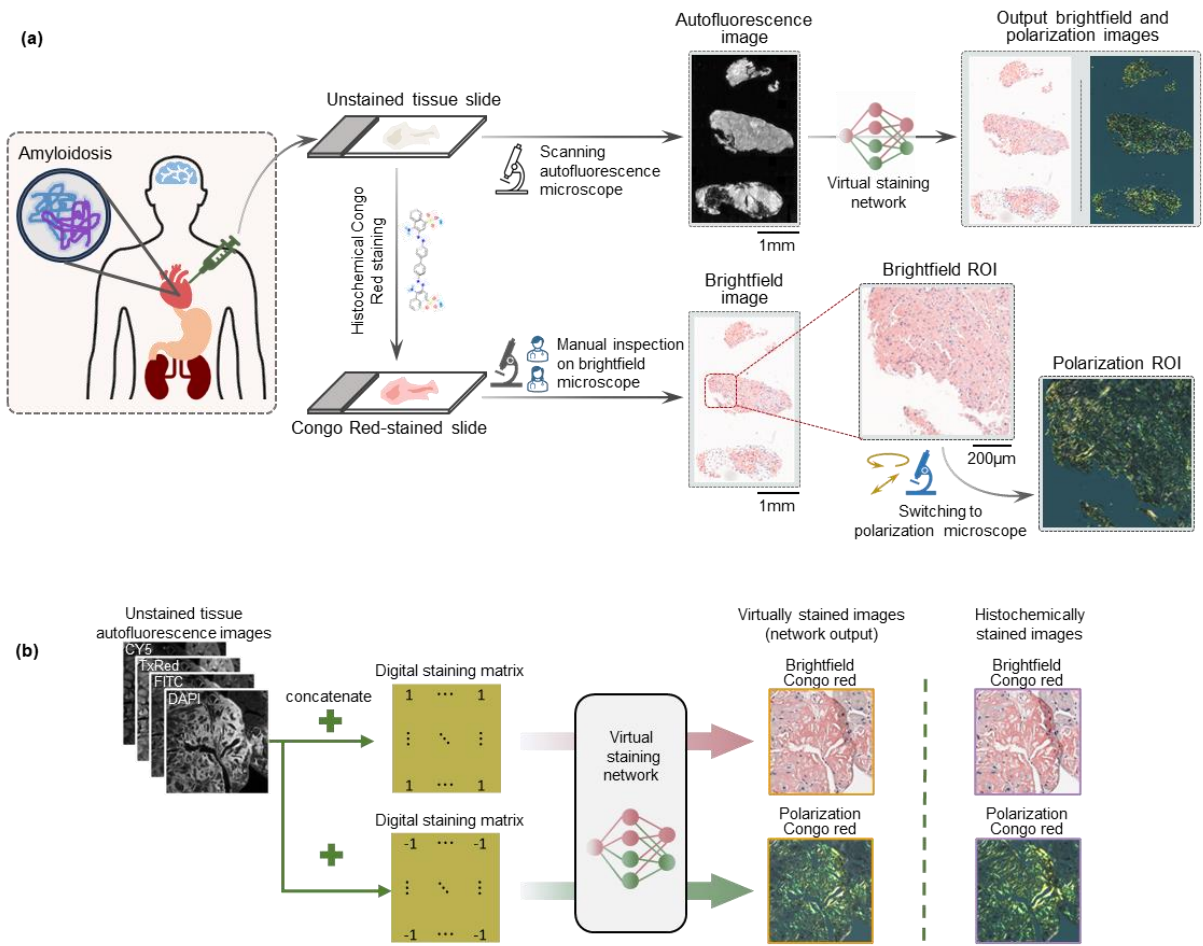

**Figure 1**. (a) Virtual birefringence imaging and histological staining of amyloid deposits in label-free tissue and its comparison to clinical workflow. (b) Virtual staining network and digital staining matrix framework to generate two output modalities (brightfield and birefringence channels).

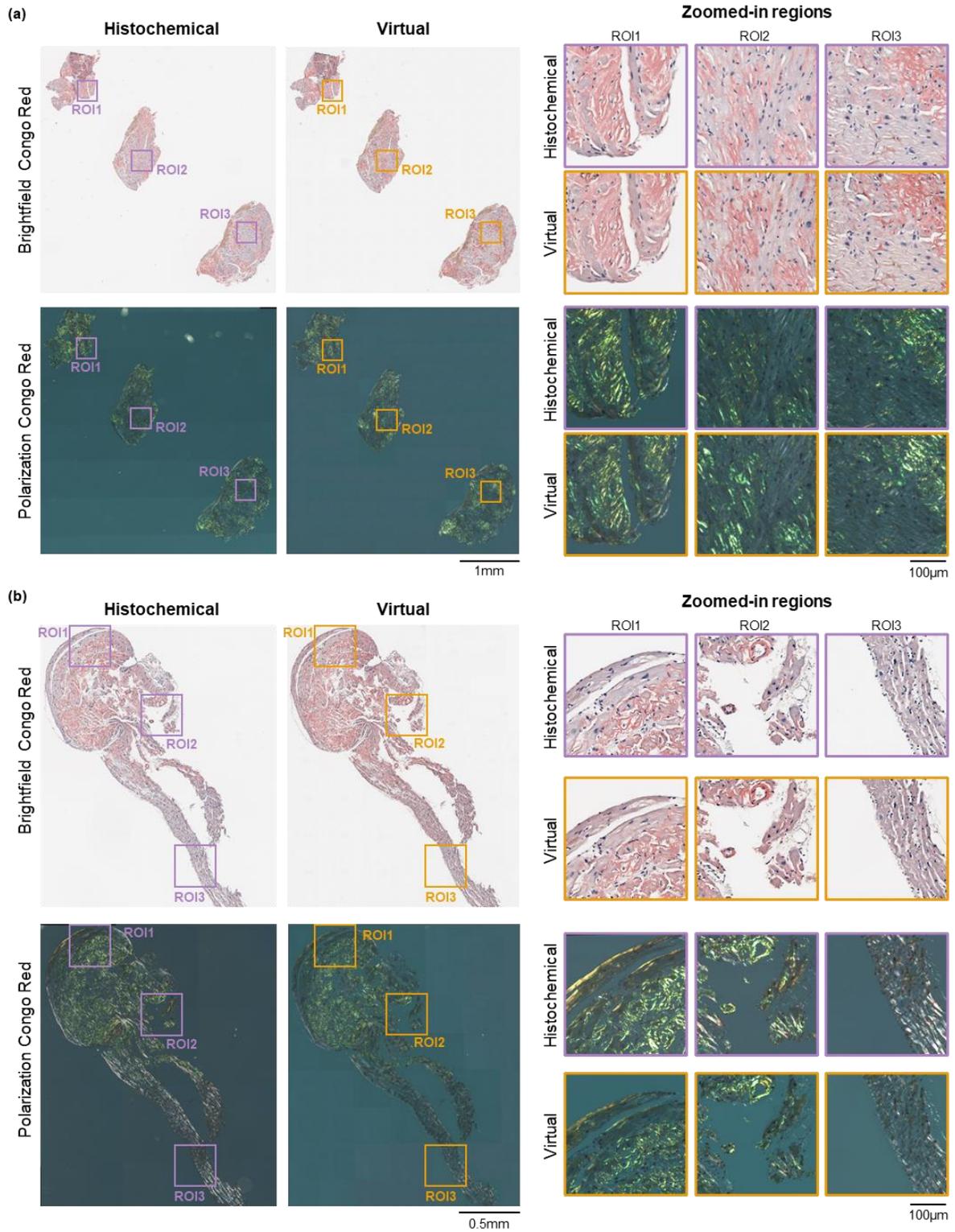

**Figure 2**. Blind testing results of virtual birefringence imaging and histological staining of amyloid deposits in label-free whole slide images with zoomed-in regions also shown.

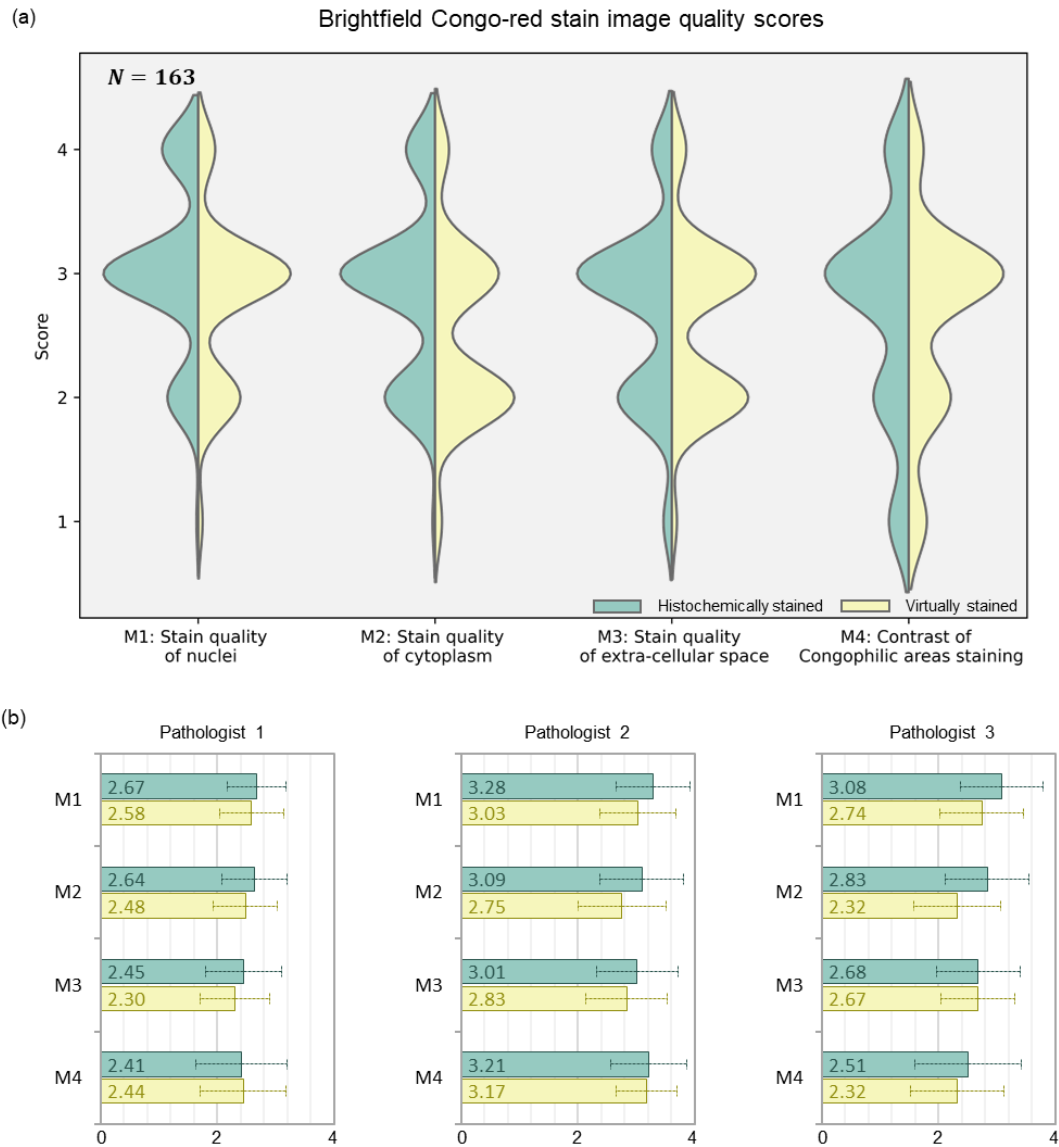

**Figure 3**. Pathologists' blind evaluation of brightfield Congo-red stain image quality (virtually stained vs. histochemically stained).

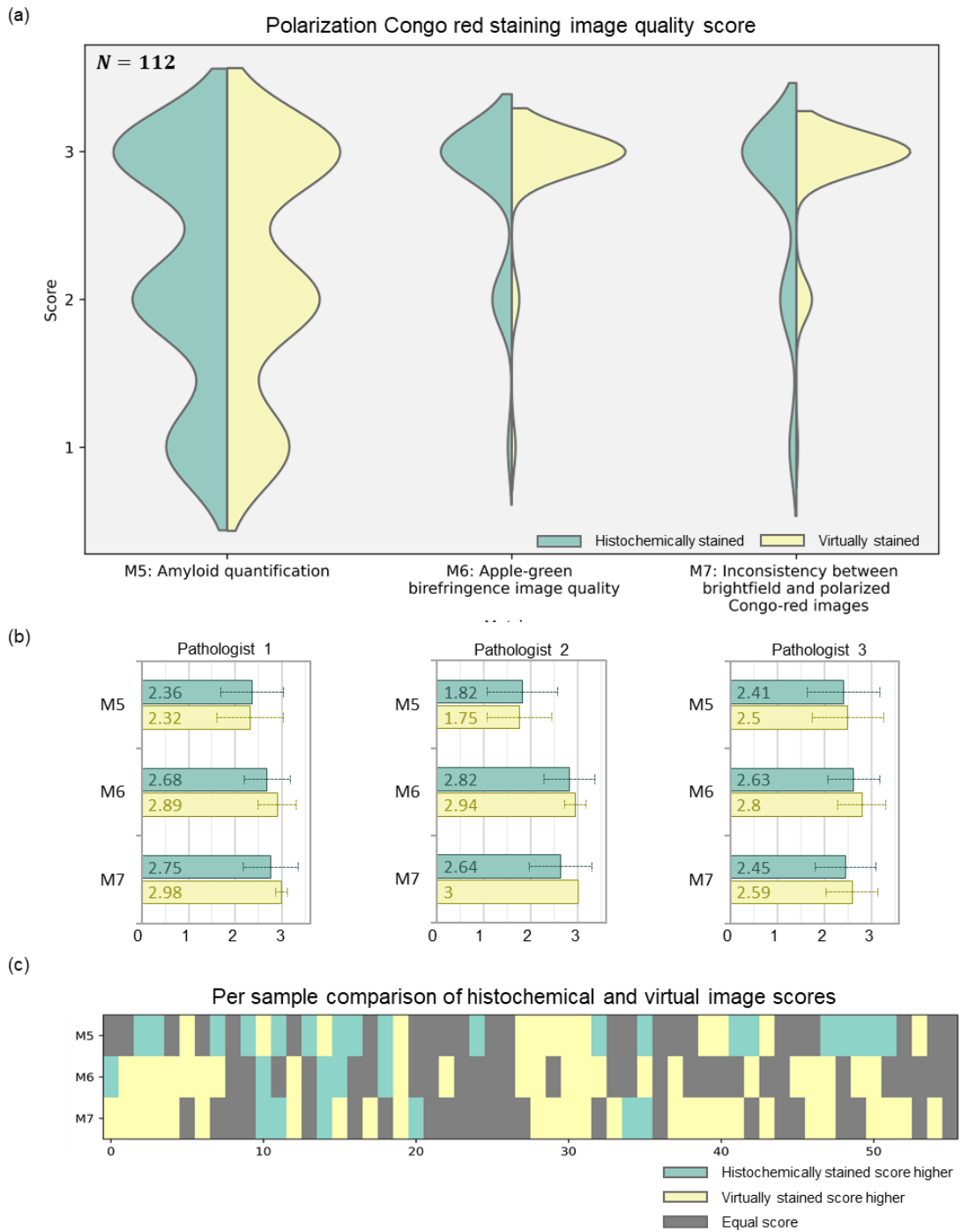

**Figure 4**. Pathologists' blind evaluation of birefringence images of Congo red staining (virtually stained vs. histochemically stained).

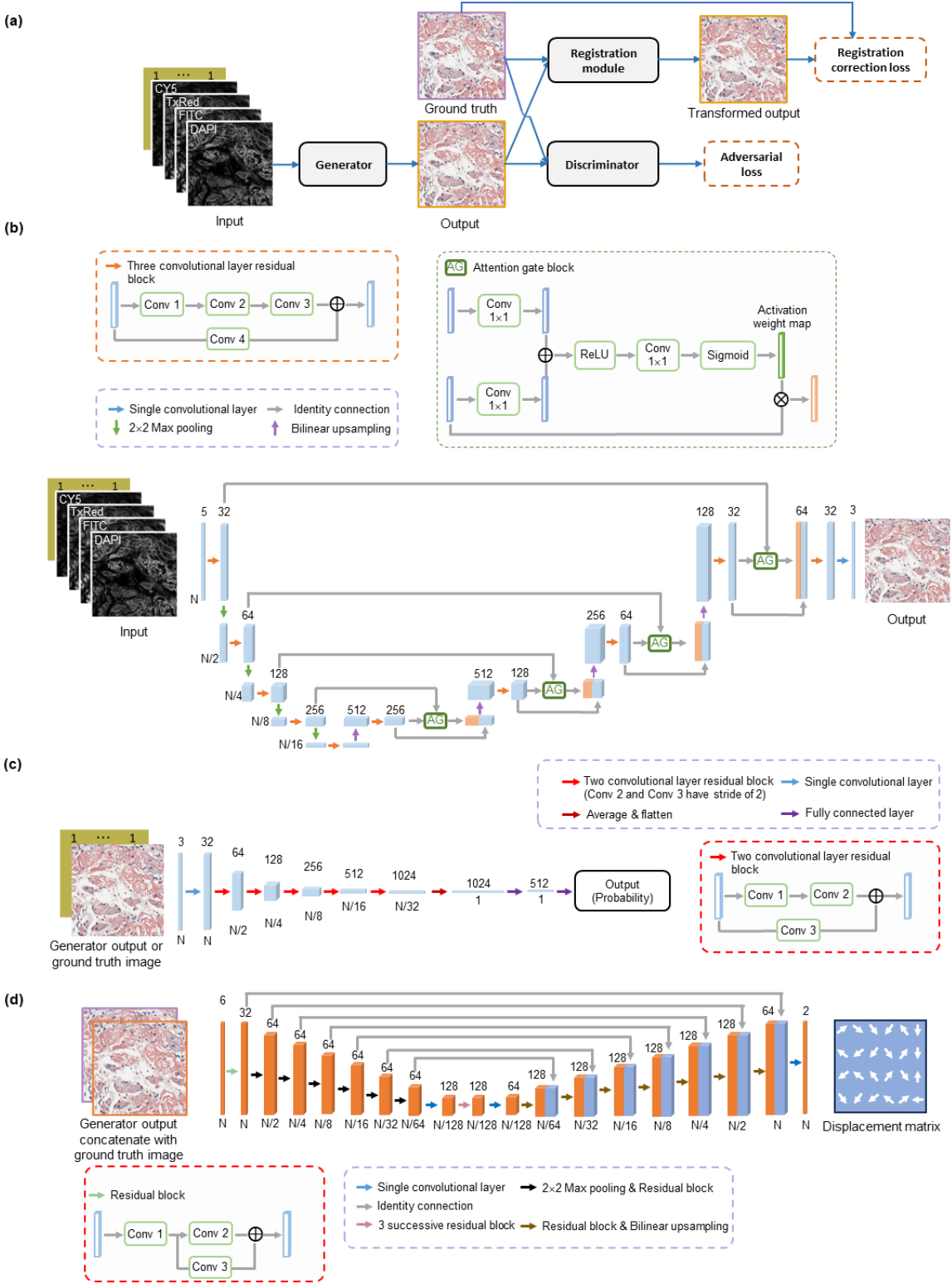

**Figure 5**. Network architecture for virtual birefringence imaging and virtual staining of amyloid deposits in label-free tissue using autofluorescence microscopy and deep learning.